\documentclass[aps,12pt,floatfix,tightenlines,showkeys,showpacs,nofootinbib,preprint]{revtex4}
\usepackage{epsfig}
\usepackage{graphicx}
\usepackage{amssymb}
\usepackage{color}
\usepackage{slashed}
\def\kp{k_\perp}\def\calM{{\cal M}}
\def\p{\phi}

\newcommand{\la}{\langle}
\newcommand{\ra}{\rangle}

\newcommand{\ben}{\begin{displaymath}}
\newcommand{\een}{\end{displaymath}}
\newcommand{\be}{\begin{equation}}
\newcommand{\ee}{\end{equation}}
\newcommand{\bea}{\begin{eqnarray}}
\newcommand{\eea}{\end{eqnarray}}

\newcommand{\eq}[1]{Eq.~(\ref{#1})}

\def\g{\gamma}
\def\e{\epsilon}

\def\a{\alpha}
 \def\d{\delta}

\def\n{\nu}
\def\e{\epsilon}
\def\L{\Lambda}
\def\G{\Gamma}

\def\s{\sigma}
\def\S{\Sigma}

\newcommand{\fs}[1]{\slashed{#1}}
\begin{document}

\preprint{NT@UW-16-03}                                                                           
\title{\bf % \hskip10cm NT@UW-10-??\\
{Meaning of the nuclear wave function}}

\author{  John D. Terry$^{1,2}$,  Gerald A. Miller$^2$}

\affiliation{$^1$Department of Physics, Univ. of California Santa Barbara, Santa Barbara CA 93106-9530 }

\affiliation{$^2$Department of Physics,
University of Washington, Seattle, WA 98195-1560}

\date{\today}

\begin{abstract}

{\begin{description}
\item[Background] The intense current experimental interest in studying the structure of the deuteron and using it to enable accurate studies of neutron structure motivate us to examine the four-dimensional  space-time  nature of the nuclear   wave function, and the various approximations used to reduce it to an object that depends only on  three spatial variables. 
\item[Purpose]  The aim  is to determine if  the ability to understand and analyze measured experimental cross sections is  compromised by making  the reduction from four to three dimensions. 
 \item[Method]   Simple, exactly-calculable, covariant  models  of a bound-state wave state wave function (a scalar  boson made of two constituent-scalar bosons)
  with parameters chosen to represent a deuteron are  used to 
 investigate the accuracy  of using different approximations to the nuclear wave function to compute the quasi-elastic scattering cross section. Four different versions of the wave function are defined (light-front spectator, light-front, light-front with scaling and non-relativistic) and used to compute the cross sections as a function of
how far off the mass-shell (how virtual) is  the struck constituent.
  \item[Results]  We show that making an exact calculation of the quasi-elastic scattering cross section involves using the light-front spectator wave function. All of the other approaches fail to reproduce the model exact calculation   if the value of Bjorken $x$ differs from unity.  The model is extended to consider an  essential  effect of spin to show that 
constituent nucleons cannot be treated as being on their mass shell even when taking the matrix element of a `good'  current.
\item[Conclusions]  It is necessary to develop realistic light-front spectator wave functions to meet the needs of current and planned experiments.  \end{description}}
\end{abstract}\pacs{nn??}

\maketitle     
\noindent
\section{Introduction}
Nuclear theorists have made  tremendous progress during the last decade  in computing nuclear spectra  from first principles~\cite{Pudliner:1997ck,Navratil:2000ww,Epelbaum:2013paa}. Two and three- nucleon interactions, with a traceable connection to QCD~\cite{Beane:2001bc,Bedaque:2002mn,Epelbaum:2008ga,Machleidt:2011zz}, have been used in exact calculations of the energy levels of nuclei with $A\le12$. Furthermore, a variety of new techniques to treat heavy nuclei have been developed. Nevertheless, some fundamental questions regarding the nature of the nuclear wave function remain.  

The nuclear wave function depends on only three of the four available space-time variables. The usual derivation of three-dimensional physics  starts with the four-dimensional  Bethe-Salpeter equation for two nucleons, which in principle makes a  non-perturbative sum of  the effects of  all interactions, and reduces it to a three-dimensional equation   without changing the  unitarity properties.  The result of the procedure is that the square of the four-momentum of the nucleons is equal to the mass squared; the nucleons are placed on their mass-shell~\cite{Blankenbecler:1965gx}. If the relativistic phase-space factor   is replaced by the non-relativistic version, the resulting equation is the Lippmann-Schwinger LS equation, equivalent to the Schroedinger equation.  The two-nucleon potential, as constrained by phase shifts computed within the LS  equation, is then used   to compute nuclear properties, by solving or approximating the many-body Schroedinger equation. Three-nucleon forces are also included.
Within this procedure the  constituent nucleons of nuclear wave functions are on
 their mass-shell. However, the sum of their basis-dependent single-particle energies
is not the energy of the nucleus; the nucleons are therefore termed as being off the energy shell.

Carrying our the reduction from four to three dimensions can be effected using either the standard equal time formulation in which the relative time  is set to 0, or
 the light-front procedure in which the relative value of $z+ct$ is set to 0~\cite{Frankfurt:1981mk,Miller:1997cr,Miller:1999ap,Miller:2009fc}. 
An exception to this procedure is the use of the Gross   equations~\cite{Gross:1969rv,Buck:1979ff,Gross:1989qj,Stadler:1996ut}, rooted in atomic physics~\cite{Eides:2000xc}, that places only one nucleon (the `spectator') on its mass shell.  No applications of this procedure to nuclei with $A>3$ exist at this time.

The purpose of the present manuscript  is to examine and determine the limitations of the three-dimensional approach to the nuclear wave function through exact and approximate evaluations of quasi-elastic scattering on a two-body system which is a semi-realistic, but completely  Lorentz-invariant  version of the deuteron. 
We concentrate on the deuteron because it is the simplest nucleus, and because there is now intense experimental interest in a variety of measurements that focus
on its wave function. For example, there is much attention on studying the wave function at high momentum transfer~\cite{Gilman:2001yh,Arrington:2011xs,Boeglin:2015cha}.   Another experiment of high interest is the proposed measurement  of
$A_{zz} $ (JLab LOI12-14-002), available by  using a tensor polarized deuteron target, aimed specifically  at studying the deuteron wave function~\cite{Long:2014fda}.
 In the quasielastic region, $A_{zz} $ can be used to compare light cone calculations with  calculations that incorporate the virtual nature of the struck nucleon, and is an important quantity to determine for understanding tensor effects. Such effects are related to the  dominance of $pn$ correlations in nuclei
 \cite{Tang:2002ww,Piasetzky:2006ai,Shneor:2007tu,Weinstein:2010rt,Hen:2014lia}. 
 The measurements are planned to occur at values of Bjorken $x={Q^2\over 2m_N \n}$ significantly greater than unity.
  Furthermore, the light-front deuteron wave function is 
 needed to interpret existing and planned  spectator-tagging experiments~\cite{Baillie:2011za,Cosyn:2014zfa}
aimed at determining neutron structure.

The  outline of the remainder of this paper follows. Our simple model is defined  in Sec.~II. The deuteron is treated as  a scalar boson that is a bound state of two different scalar bosons. The vertex function is taken as a constant. Such models have long been used~\cite{Gunion:1973ex,Miller:2009sg}  to illustrate relativistic aspects of complicated dynamical situations. 
The  definition of  constituent  virtuality is presented and its importance is illustrated in Sec.~III.  The model exact calculation of the analog of quasi-elastic electron-deuteron scattering cross section is presented in Sec.~IV. We neglect the influence of final state interactions throughout this paper. This simplification allows us to focus on the influence  of virtuality. Furthermore, the effects of final state interactions can be minimized through the appropriate choice of kinematics~\cite{Boeglin:2015cha}.  A discussion of four different wave functions: light-front-spectator, light front, light front with Bjorken scaling, and non-relativistic is presented in Sec.~V.  Cross sections obtained using these different models are compared with the model exact cross sections in Sec.~VI. All of the models, except the light-front-spectator,  fail badly if the value of $x$ differs significantly from unity.  The vertex function of the simple model of Sect.~II is generalized in Sect.~VII, where it is shown that the qualitative conclusion just stated does not depend on using a constant vertex function. 
One aspect of spin is considered in Sec.~VIII where it is shown that it is necessary to consider the virtual nature of constituent fermions, even  if computing the matrix element of a `good' current. The final section presents a summary and discussion of the possible implications of the work presented here. 
% However, the numerical examples presented here are new, the 
%current relevance, as driven by experimental needs, is very high.

\section{Model dynamics and model scattering process}

The nuclear dynamics are modeled by a version of the $\p^3$ model, generalized to $D\p\chi$ so that one scalar particle a `deuteron'  $D$ of mass $M$ is a bound system of two different scalar particles $\p,\chi$ of mass $m$, only one of which interacts with a scalar probe of four momentum $q$. The interaction between the probe and the   struck nucleon is  taken to be  a constant, $g$. The  deuteron vertex function $\G(k,P)$  is also taken as a constant, $G$.
This set of  dynamics corresponds to the  0'th order chiral perturbation
theory version of the deuteron. The model allows for all matrix elements to be computed in covariant fashion and there is no need to limit the kinematics. The values of $m,M$ are those of the average nucleon mass and mass of the physical deuteron. Thus $M=2m-B$, with $B=0.0022 $ GeV. These scales $B$ and $M$ span the range of mass scales that would enter into a more realistic model.

The quasi-elastic scattering reaction of interest is shown in Fig.~1. A scalar `deuteron'  of 4-momentum $P$ encounters a virtual space-like scalar  photon of four-momentum $q$, leading to a final state in  which the struck nucleon has momentum $k+q$, and is a real particle of positive energy. The spectator $ s$ has four-momentum $p_s$ with $p_s^2=m^2$ and its energy is greater than 0.  We consider only this diagram here so as to concentrate on the fundamental aspects. Therefore the usually important effects of final state interactions are neglected throughout this paper, 
%\end{document}
\begin{figure}[h]\label{diag}
\centering
\includegraphics[scale = 0.25]{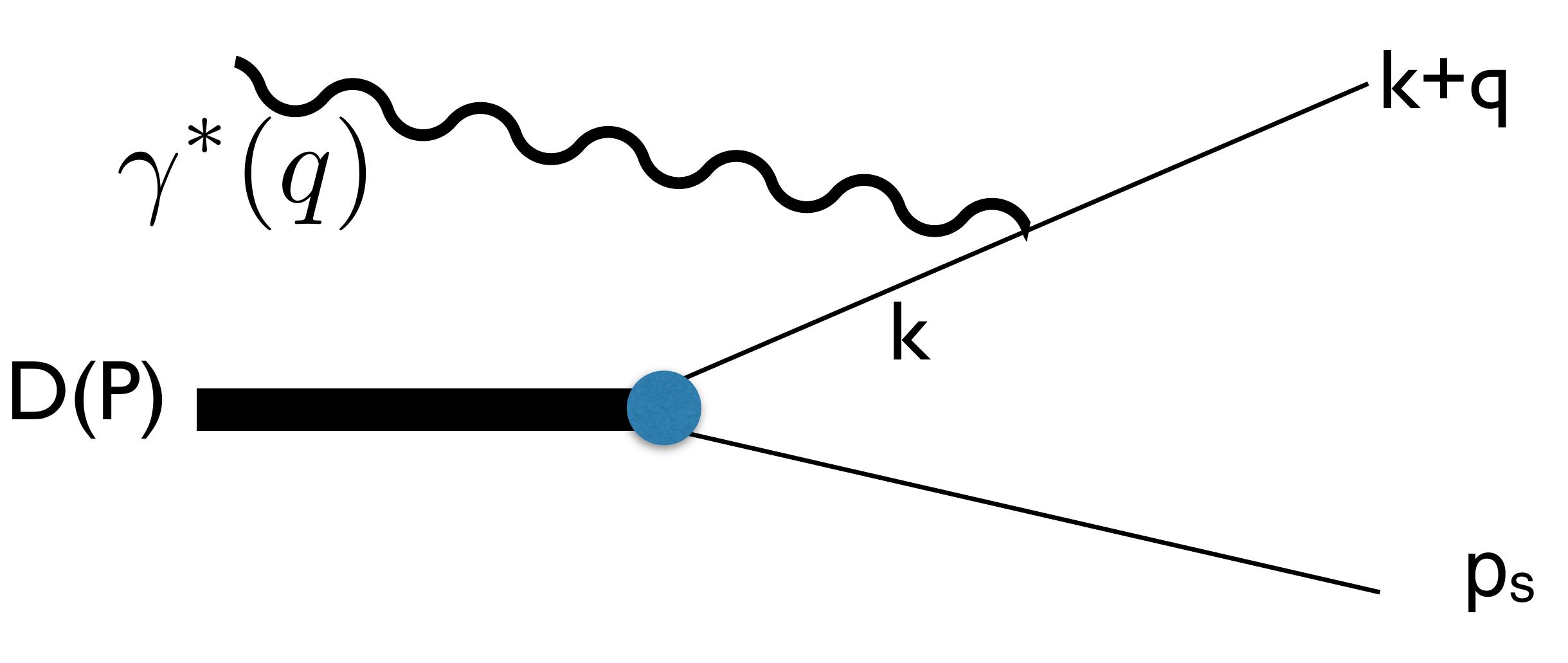}
\caption{(Color online) Scattering process of interest}
\end{figure}
%\section{Light Front Wavefunction}
We use the convention that in the deuteron rest-frame the  four-momentum $q$ is given by $q=(\n,0,0,-\sqrt{Q^2+\n^2}),$  with $Q^2=-q^2>0.$
It is useful to define two kinematic variables
\bea x\equiv {Q^2\over 2m\n},\eea which ranges between 0 and  about $M/m\approx 2$, and 
\bea \xi \equiv  {Q^2\over Mq^-}=-{q^+\over M}={m\over M}{2x\over 1+\sqrt{1+{4m^2x^2\over Q^2}}}={m\over M}{2xQ\over Q+\sqrt{Q^2+{4m^2x^2}}},\label{xi}\eea  $\xi$ is the Nachtmann variable for the given target, and it is limited by momentum conservation to be 
less than unity. The $\pm$ components of the four-momentum $V$ of any particle is defined here as $V^\pm\equiv V^0+V^3$.  Another useful variable is
the light-front variable $\a$, defined via
\bea p_s^+\equiv (1-\a) P^+=E_s+p_{sz}.\label{al}\eea
In the deuteron rest frame $P^\pm=M$.
\section{Nucleon Virtuality}
The virtual nucleon in Fig.~1 has four momentum $k$ given by $k=P-p_s$. The quantity \bea V\equiv m^2-(P-p_s)^2=m^2-k^2 \equiv-X\eea measures the deviation of the about to be struck nucleon from its mass shell. The founding assumption of nearly all nuclear wave functions is that the virtuality vanishes. 
%Its inverse is the Bethe-Salpeter wave function of the nucleus if the vertex function is constant.

Proceed by examining the quantity $X=-V$. %This is facilitated by defining  via
 %where the superscript plus denotes that we take the $+$-component of the four momentum.
Given that $k^-=P^--p_s^-$,  the use of \eq{al} gives  $k^+=\a M,\,$  in the deuteron rest frame, and also $k^-=P^--p_s^-=M-{\kp^2+m^2\over (1-\a)M},$ where $\bf \kp $ is the momentum of the spectator nucleon.
Furthermore
\bea
X(\a,\kp)&=(P^+-p_s^+)(P^--p_s^-)- k_\perp^2-m^2 %=\a M(M-{ k_\perp^2+m^2\over (1-\a)M})- k_\perp^2-m^2\nonumber\\
%=\a M^2-{ k_\perp^2+m^2\over 1-\a}
=\a(M^2-{ k_\perp^2+m^2\over \a(1-\a)}).\label{lf00}
\eea
We examine the delta function 
%\bea
$\d( X+2k\cdot q-Q^2)$  %\,\label{df}\eea
 to determine the relevant value of $\a$ as a function of $q$ and $\kp$:

\bea
%X+q^+k^-+q^-k^++q^-q^+=0\\
&X+q^+(M-{ k_\perp^2+m^2\over M( 1-\a)})+q^- \a M-Q^2=0\label{X0}.
\eea
%\a(M^2-{ k_\perp^2+m^2\over \a(1-\a)})+q^+(M-{ k_\perp^2+m^2\over M( 1-\a)})+q^-\a M-Q^2=0\\&
%(\a-\xi)M^2-{ k_\perp^2+m^2\over (1-\a)}-q^+{ k_\perp^2+m^2\over M(1-\a)}+q^-M(\a-\xi)=0\\&
%(\a-\xi)M^2-{ k_\perp^2+m^2\over (1-\a)}-q^+{ k_\perp^2+m^2\over M(1-\a)}+q^-M(\a-\xi)=0\\&
The vanishing of the effect of the  virtuality at high momentum transfer and energy  can most readily be observed by using light-front variables.
The argument of the delta function shown in \eq{X0} can be rewritten as
\bea&(\a-\xi)(q^-M+M^2)-(1-\xi){ k_\perp^2+m^2\over (1-\a)}=0.
\label{ex} \eea
It is worthwhile to point out that in the Bjorken limit of $Q^2/\n^2\ll1,\,$ $q^+\ll q^-$,  the last two   terms of \eq{X0} are much larger than the first two terms. In that case, one may ignore  the 
the  first two terms, so that in this scaling limit $\a=\xi$. 
%If the first two terms of \eq{X0} are ignored, one sees that $\xi=\a$, which corresponds to the parton model.

We need to examine the effects of the first two terms of \eq{ex}. This  is a quadratic equation in $\a$, which can be solved, yielding   %There are two real roots. Only the one with $1-\a>0$ leads to a positive value of $E_s$.
the result  
\bea \a={1\over2}\left(1+\xi-\sqrt{(1-\xi)^2-\e{4C(1-\xi)}}\right),\label{a}\\
%\frac{\xi-\e+\sqrt{(\xi-\e)^2+4C\e (1-\e)}}{2(1-\e)},\, \\
\e\equiv {M\over M+q^-}<1,\,C\equiv {( k_\perp^2+m^2)\over M^2} \label{eqa}
\eea
obtained using the condition that when $\e C=0,\, \a=\xi$, and
$ k_\perp$ is the perp component momentum of the spectator. % In the limit that $\e\to0$ we find $\a\approx \xi$ (parton model). Otherwise $\a$ depends on  $ k_\perp$.
%The argument of the square root must be positive, otherwise there is no solution that conserves energy for the   parameters $\xi,C,\e$. If $\e<1$,
The relations \eq{a} and \eq{xi} determine the spectator momentum for each value of $\kp$.

The value of $\a$ must be such that the spectator energy $E_s\ge m$.
This means that
\bea {1\over2}[M(1-\a)+{k_\perp^2+m^2\over (1-\a)M}]\ge m\label{cond}\eea
 Note that $M=2m-B,\,B>0.$
\eq{cond} holds for all values of $\a$ such that  
$  \a<1 .$
 Conservation of four-momentum leads to a limit on $x\equiv {Q^2\over 2 m \n}:$
 \bea x\le {M\over m} {1\over 1+{4m^2-M^2+4k_\perp^2\over Q^2}}.
 \eea
 This equation can also be written as a limit on $k_\perp^2:$
 \bea Q^2+4m^2-M^2+4k_\perp^2<2M\n
 .\eea

 Armed with the value of $\a$ which depends on $x,Q^2$ and $\kp$, we may compute the value of $V$ for different kinematic situations. The results are shown in Figs.~2, 3 and 4.
 \begin{figure}[h]\label{V1}
\centering
\includegraphics[scale = 0.6]{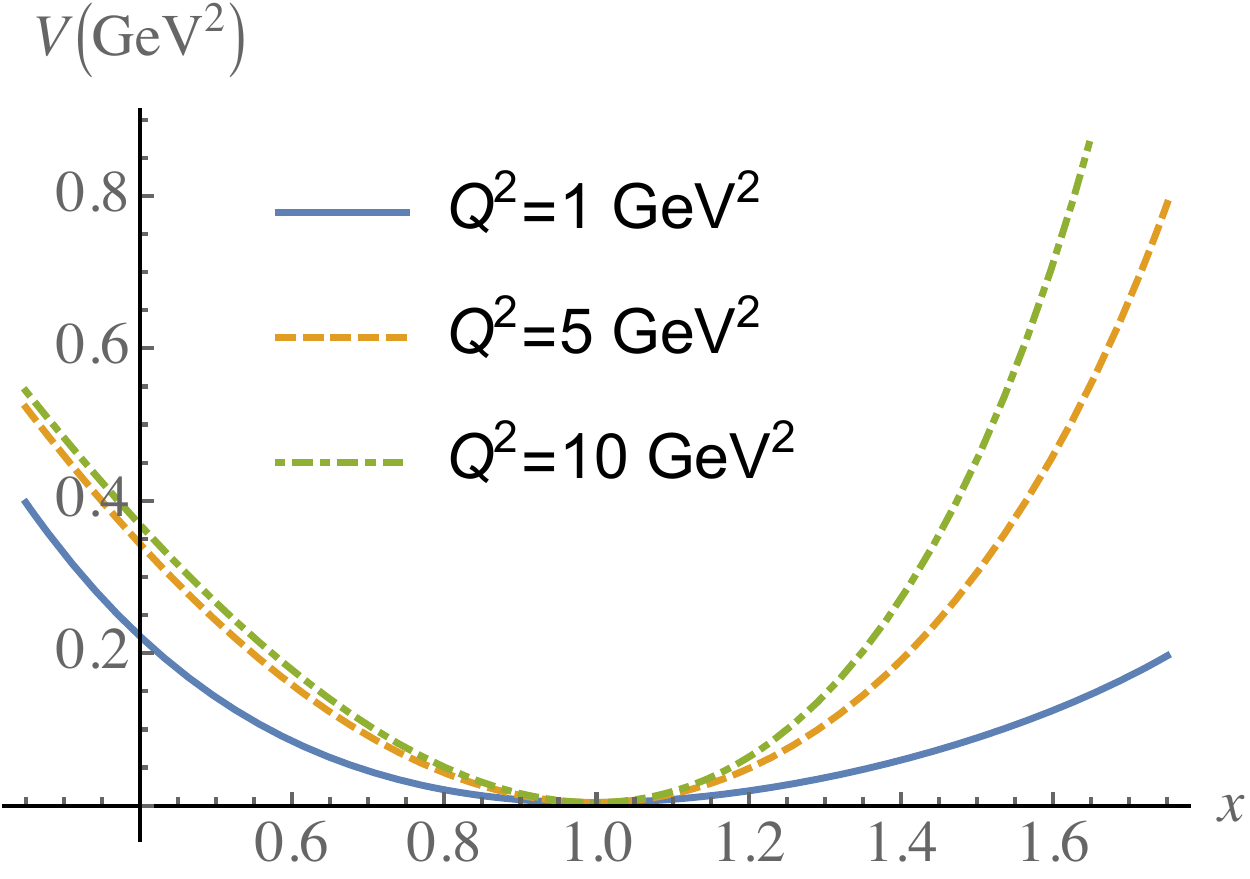}
\caption{(Color online) Virtuality as a function of $x,Q^2$ for $\kp=0$.}
\end{figure}
\begin{figure}[h]\label{V2}
\centering
\includegraphics[scale = 0.6]{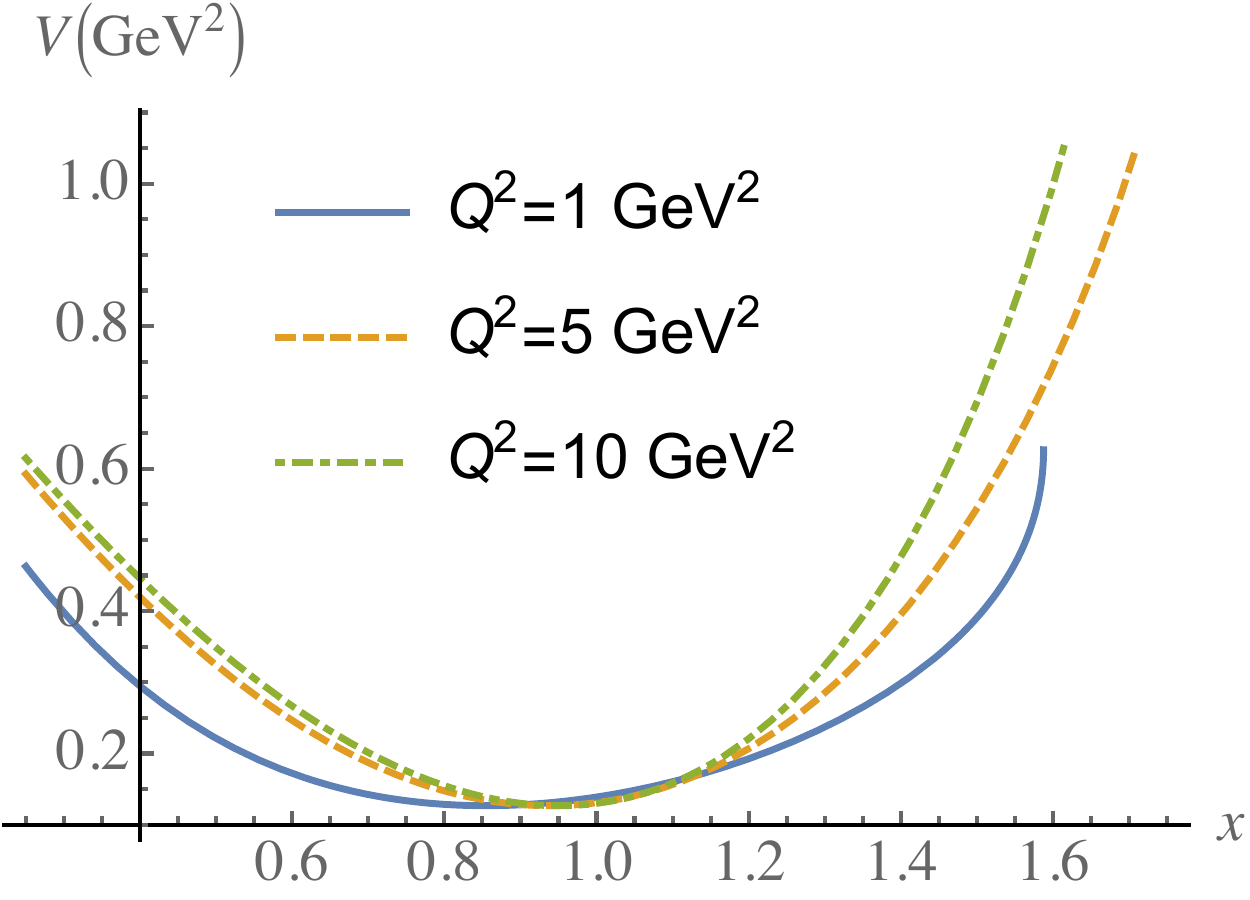}
\caption{(Color online) Virtuality as a function of $x,Q^2$ for $\kp=0.25$ GeV.}
\end{figure}
\begin{figure}[h]\label{V4}
\centering
\includegraphics[scale = 0.6]{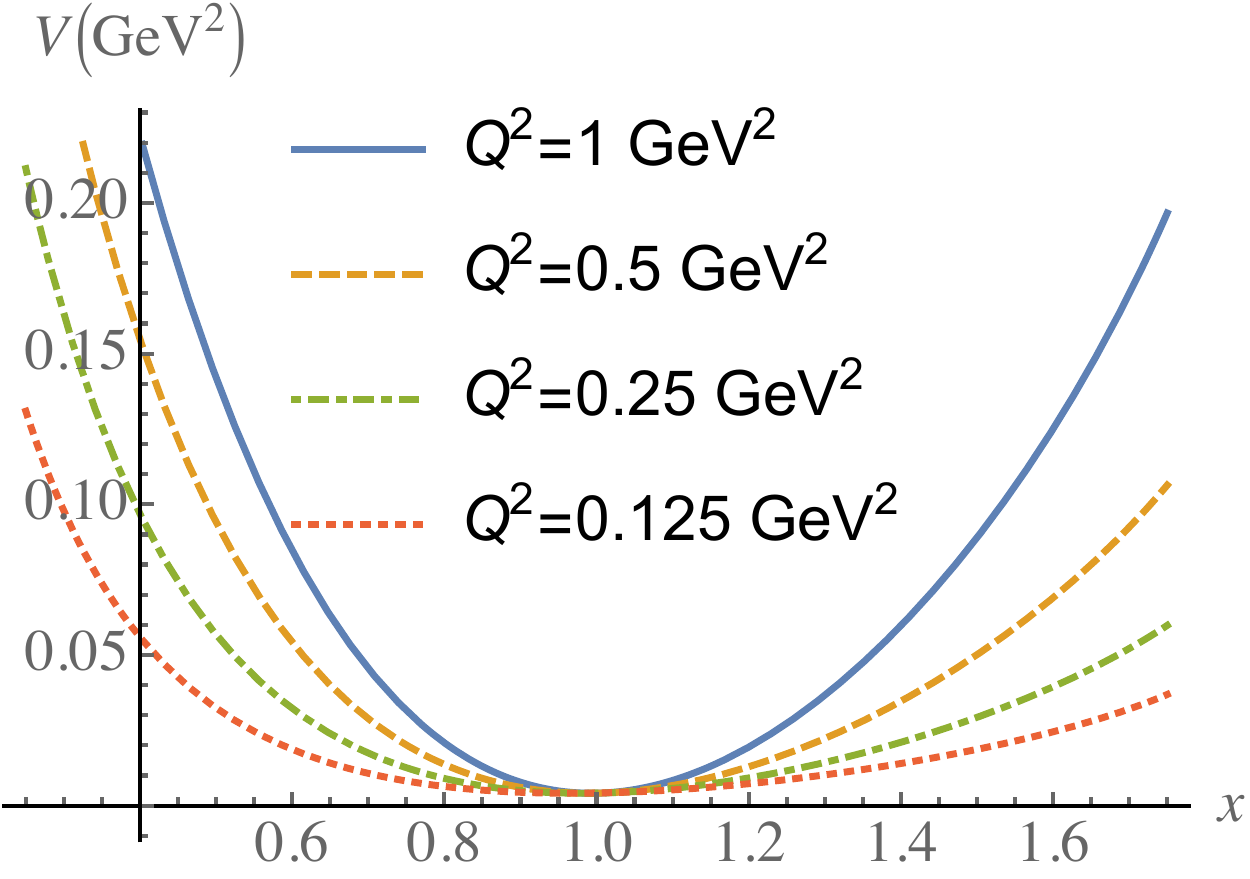}
\caption{(Color online) Virtuality as a function of $x,Q^2$, lower values of $Q^2$ for $\kp=0$.}
\end{figure}
We see that, except for  small values of $Q^2$ and  $x$ near unity,  the value of $V$ is generally larger than $0.1$ GeV$^2$. This corresponds to a momentum of 300 MeV/c, which is not an ignorable scale in nuclear physics. Thus  in general
 approximating $V$ by 0 is expected to be a dangerous approximation. This means that the usual nuclear procedure of treating the nucleons as being on their mass shell is not valid, and that  the connection between the scattering amplitude and the usual  equal-time or light front three-dimensional wave functions is severed.
\section{exact  model quasi-elastic  cross section}
The cross section is for the absorption of a space-like  scalar ``photon" of four-momentum $q$ on a two body system of scalar mesons which is our toy model of the deuteron.
In this case
\bea 
d\s ={(2\pi)^4\over j} {d^4p\over(2\pi)^3}\d_+(p^2-m^2) {d^3p_s\over 2E_s(2\pi)^3} \delta^4(P+q-p-p_s)|\calM|^2, \label{sig}
\eea
where $\calM$ is the invariant amplitude, and $j=4M|\vec{q}|$ is the flux factor. Define $k=P-p_s$, the four-momentum of the struck particle.
Anticipating the use of light-front variables, we state
\bea {d^3p_s\over 2E_s}={d^2\kp dk^+\over 2(P^+-k^+)},\eea
so that integration over the four-momentum-conserving delta function yields
\bea 
d\s =\d_+((k+q)^2-m^2) {d^2\kp dk^+\over 2(P^+-k^+)(2\pi)^2j}|\calM|^2.
\eea
For our model 
\bea |\calM|^2={g^2G^2\over X(\a,\kp)^2},\eea
where $X(\a,\kp)$ is the absolute value of the inverse propagator. Then
%Here the final state $f$ is just the spectator nucleon. 
we write 
\bea d\s={d^2k_\perp\over 8\pi^2j} \int{dk^+\over p_i^+- k^+}\delta_+((k+q)^2-m^2) {g^2G^2\over X(\a,\kp)^2}
\eea
  The value of $X(\a,\kp)$ is  given by \eq{lf00}.
  %fixed by the delta function of  \eq{lf} to be 
%\bea
%-X=2k\cdot q-Q^2=\a Mq^-+(M-{k_\perp^2+m^2\over (1-\a)M})q^+-Q^2=(\a-\xi)Mq^--\xi(M^2-{k_\perp^2+m^2\over (1-\a) })\label{X1}
%\eea
We do the integral over $dk^+$ to obtain
\bea j {8\pi^2\over g^2G^2}{d\s\over d^2k_\perp}= %\Theta(E-m){1\over( 1- \a\,)X^2}{1\over q^-M+M^2}{1-\a\over 1-2\a+\xi},\label{31}\\=
 %\Theta(E-m)
 {1\over X(\a,\kp)^2}
 {1\over q^-M+M^2}{1\over 1-2\a+\xi},\label{dsig}\eea
where $E=\nu+M-E_s$ and 
$\a$ as given by \eq{a} and  $X$ of \eq{lf00} are functions of $\n,Q^2,k_\perp^2$. 
Note that the limit $\a\le1$ is enforced by \eq{a}, and the limit $\xi\le1$.
%The last term in \eq{31} comes from the evaluation of the delta function.
The use of light-front variables  is not necessary, but  their use does  simplify the evaluation. We have obtained equivalent results using the standard energy-momentum variables.

It is convenient to define the quantity
\bea {d\Sigma\over d^2k}\equiv  j{8\pi^2\over g^2G^2}{\n d\s\over  d^2\kp}\eea
The factor $\n$ is  inserted because  the cross section for a single free nucleon can be interpreted to have this  factor. Thus  ${d\Sigma\over d^2k}$  represents  a cross section per nucleon.
Results for   ${ d\S\over  d^2\kp}$  are shown for two  ranges of $Q^2$ in Fig.~5. We begin by noting that  the cross sections look qualitatively similar to measured experimental data (see {\textit e.g.} Fig.~6 of Ref.~\cite{Boeglin:2015cha})  giving some credence to the simple model we use. Note also that the scaling limit (in which the cross sections depend on $x$ but not on $Q^2$)  is obtained for $Q^2$ of order 10's of GeV$^2$.

\begin{figure}[h]\label{dse}
\centering
\includegraphics[scale = .7]{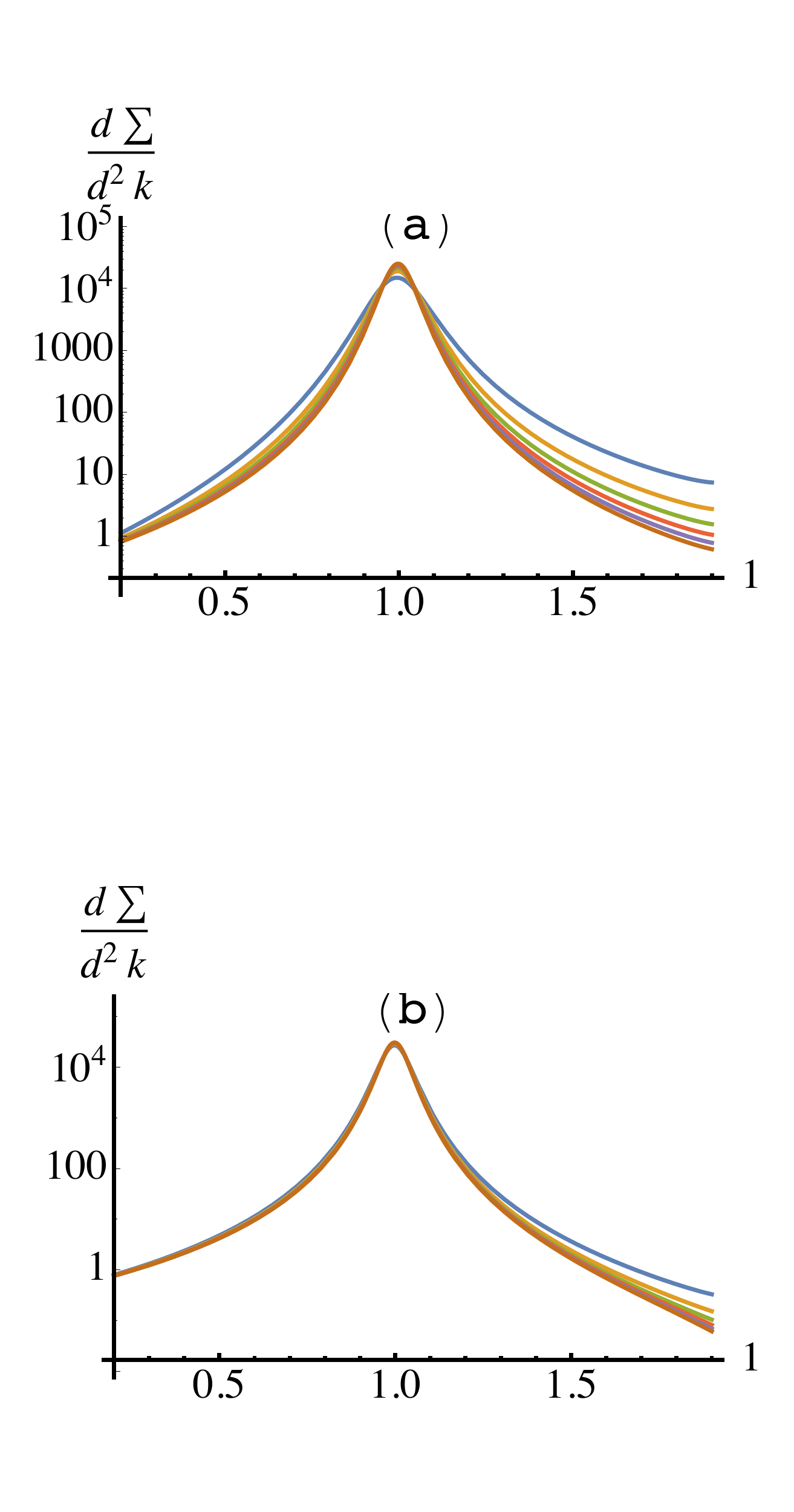}
\caption{(Color online) ${d\S\over  d^2\kp}$ as a function of $x$ for two different ranges of $Q^2$ with  $\kp=0$. (a) $Q^2$ from 1 to 6 GeV$^2$. (b)  $Q^2$ from 10 to 60 GeV$^2$. For each case, the lower the value of $Q^2$, the higher the  cross section. For larger values of $Q^2$ the curves tend to coalesce. }% Bottom: $Q^2$ from 100 to 600 GeV$^2$}
\end{figure}
\section{Wave functions}
We next relate the exact model calculations of the previous Section with various ideas about wave functions that are in the literature.

\subsection{Exact model calculation uses  the light-front spectator wave function}
The Bethe-Salpeter~\cite{Salpeter:1951sz}
wave function for this model is given by 
\begin{equation}
\Psi(k, P) = \frac{-iG}{(k^2-m^2+i\epsilon)((P-k)^2-m^2+i\epsilon)}.
\end{equation}
This quantity does not enter in the calculation of the invariant amplitude,  which  involves only $ \frac{-iG}{(k^2-m^2+i\epsilon)}=-iG/X$
However, the factor $1/X$  can be obtained by doing an integration that places the spectator particle on the 
mass shell,  so that $(P-k)^2=m^2$, with $(P^+-k^+)>0$. 
The result of this integral is the spectator wave function  of the Gross equation~\cite{Gross:1969rv} times a kinematic factor, so that 
 the object $1/X$ of \eq{lf00} corresponds to using  the spectator wave function.
Note that  the integration used here involves light front coordinates to take advantage of the high energy of the incident virtual photon.  Thus,  computation of the exact cross section makes explicit use of a light-front version of the Gross equation wave function. We may even say that the light-front spectator wave function is designed to give the correct quasi-elastic scattering cross section.
This wave function has the odd, but useful, feature that one constituent is virtual and the other spectator constituent, a spectator, is on its mass shell.

%A sample cross section is shown in Fig.~\ref{samp}
%\begin{figure}[H]
%\centering
%\includegraphics[scale = 0.5]{dsig.pdf}
%\caption{Sample cross section. $k_\perp=0$.  %he quantity $ {1\over g^2G^2}{d\s\over d^2k_\perp}$ \eq{dsig} is shown as a function of $Q^2$ in GeV$^2$ for $x= {Q^2\over 2m \n} $ ranging from 0.2 (top curve) to 3. 8 (bottom curve) The cross section vanishes for low $Q^2$ because of the constraints imposed by momentum conservation.
 %}\label{samp}
%\end{figure}
\subsection{ The on-mass shell limit uses the light front wave function}
The light front wave function is derived by taking the constituent particles to be on the mass-shell, denoted by OS. In this case $k^2-m^2=0$ and 
\eq{X0} becomes
\bea  q^+(M-{ k_\perp^2+m^2\over M( 1-\a)})+q^- \a M-Q^2=0.\label{X0os}\eea
The solution is given by 
\bea 
\a=\a_{\rm OS}=\frac{1}{2} \left(-\sqrt{\epsilon_0  \left(4 C  \xi +2  \xi ^2-2  \xi \right)+(1- \xi )^2+ \xi ^2 \epsilon_0 ^2}+ \xi  (\epsilon_0
   +1)+1\right) ,\label{aos}
   \eea
with  $ \e_0\equiv{M\over q^-}. $   The cross section is given by 
\bea j8\pi^2 d\s_{\rm OS}=d^2k_\perp \int{dk^+\over P^+-k^+}\delta_+(-\xi (M^2-{\kp^2+m^2\over 1-\a})+k^+ q^- -Q^2) {g^2G^2\over X^2(\a_{\rm OS},\kp)}.\label{OS}
\eea
The notation $X(\a_{\rm OS},\kp)$  refers to using $\a\to\a_{\rm OS}$ in the defining equation \eq{aos}.
For calculations of elastic scattering the use of the light front wave function gives the exact result~\cite{Miller:2009sg}.
%Evaluation of \eq{OS} leads to evaluating $X$ at $\a=\a_{\rm OS}$.

\subsection{ Light front wave function-with scaling }
For large value of $Q^2$ and $\n$, when $Q^2/\nu$ is constant  and $x\equiv {Q^2/2m\n}$, (the scaling limit) one may
 ignore the $k^2-m^2$ and $q^+k^-$ appearing in the argument of the  delta function of \eq{X0}. In this case 
 \bea\d( X+2k\cdot q-Q^2)=\d( k^2-m^2+2k\cdot q-Q^2)\to \d(k^+ q^--Q^2),\eea
 so that 
\bea\a=\xi ,\eea
and 
\bea  j8\pi^2 d\s_{sc}=d^2k_\perp \int{dk^+\over P^+-k^+}\delta_+(k^+ q^- -Q^2) {g^2G^2\over X(\a,\kp)^2},
\eea
% \bea
%X\to X_a=k^2-m^2= k^+k^--\kp^2-m^2=k^+(M- {k_\perp^2+m^2\over (1-\a )M})-k_\perp^2-m^2=\a (M^2 -{k_\perp^2+m^2\over\a (1-\a )})
%\eea
yielding
\bea j8\pi^2{d\s_{sc}\over g^2G^2d^2k_\perp}= {1\over (1-\xi) q^-M }\,{1\over X^2(\xi,\kp)} .%ver\a (1-\a )})^2},\,\a=\xi
\eea
 In the scaling limit, the
relevant wave function is the light front wave function evaluated at a momentum fraction $\xi$ that is determined only  by $x$ and $Q^2$. 
% In that limit, one   neglects  the virtuality (value of  $k^2-m^2$) of the struck particle that appears in the delta function 
 %because $q^-$ and $Q^2$ are assumed to be much larger than the presumed dominant scales in the problem.
 
 The net result, so far, is 
 that the exact calculation is handled by the spectator wave function. If one neglects the virtuality of the struck nucleon, one may use the light front wave function, but it is evaluated at a momentum fraction that depends upon $\kp$ as well as on $(x,Q^2)$.  Only in the scaling limit can one use the light front wave function, evaluated at the Nachtman variable $\xi$.
 
 \subsection{Non-relativistic limit}
We define the non-relativistic limit as using the non-relativistic approximation to the inverse propagator $X$ of \eq{lf00} in the expression for the cross section \eq{dsig}.
Thus \bea X=M(M-2E_s)\to -M(B+{\vec{k}^2\over m}),\eea
where in the non-relativistic approximation $E_s=\sqrt{\vec{k}^2+m^2}\approx m+{\vec{k}^2\over 2m}$. Thus   we use the non-relativistic limit and obtain 
\bea {1\over X_{NR}}={-m\over M}{1\over m B+ k_z^2+\kp^2}.\label{xnr}\eea This is essentially the non-relativistic wave function for a delta function potential, and is also the zero range wave function of Bethe~\cite{Bethe:1950jm}.

 To evaluate the cross section in \eq{dsig} we use   $X_{NR}$  of  \eq{xnr}. It is necessary to determine $k_z$ in terms of $\a$. In the non-relativstic theory all constituent particles are on their mass-shell, so $k_z$ is determined from $\a_{\rm OS} $ via
 \bea
\a_{\rm OS}={E(k)+k_z\over 2m},\,E(k)=m+{\vec{k}^2\over 2m}.\eea
Solving this equation for $k_z$ gives  the result:
\bea
{k_z\over m}=-1+\sqrt{4\a_{\rm OS}-1-{\kp^2\over m^2}},
\eea
with  $\a_{\rm OS}$ given  by \eq{aos}.
This is the root that has 
 $k_z=0$ if $\a_{\rm OS}=1/2$ and $\kp=0$.

\section{Model results and the accuracy of using  different wave functions}
\begin{figure}[h]\label{x1}
\centering
\includegraphics[scale = .5]{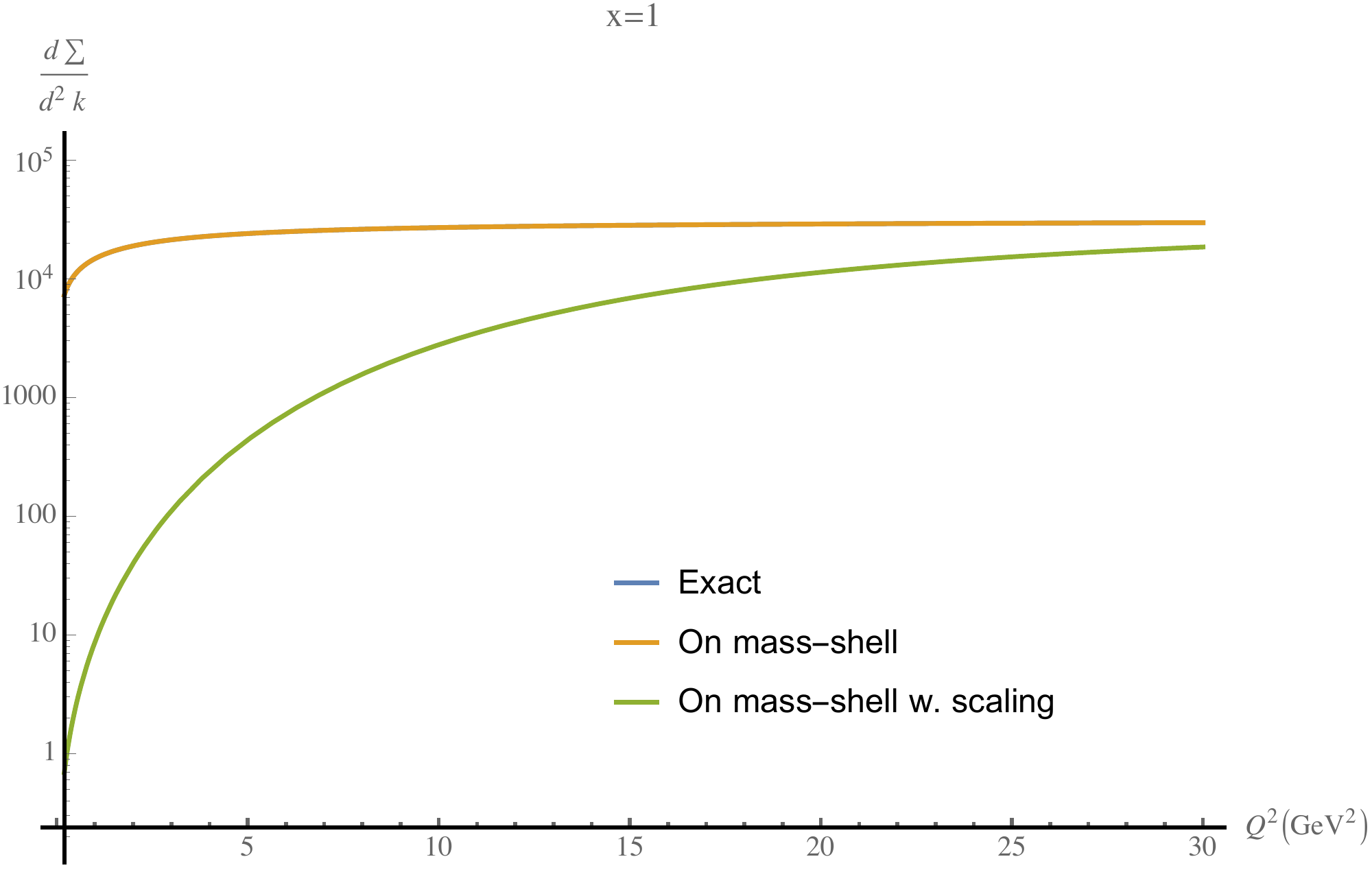}
\caption{(Color online) ${d\S\over  d^2\kp}$ for three models at $x=1,\,\kp=0$.    Only two curves are observable easily because of the confluence of the 
exact and on-mass shell light-front wave function approach. }%The lower panel shows that   the on-mass shell approximation use of the light front wave function yield .}% Bottom: $Q^2$ from 100 to 600 GeV$^2$}
\end{figure}

 \begin{figure}[h]\label{p5}
\centering
\includegraphics[scale = .5]{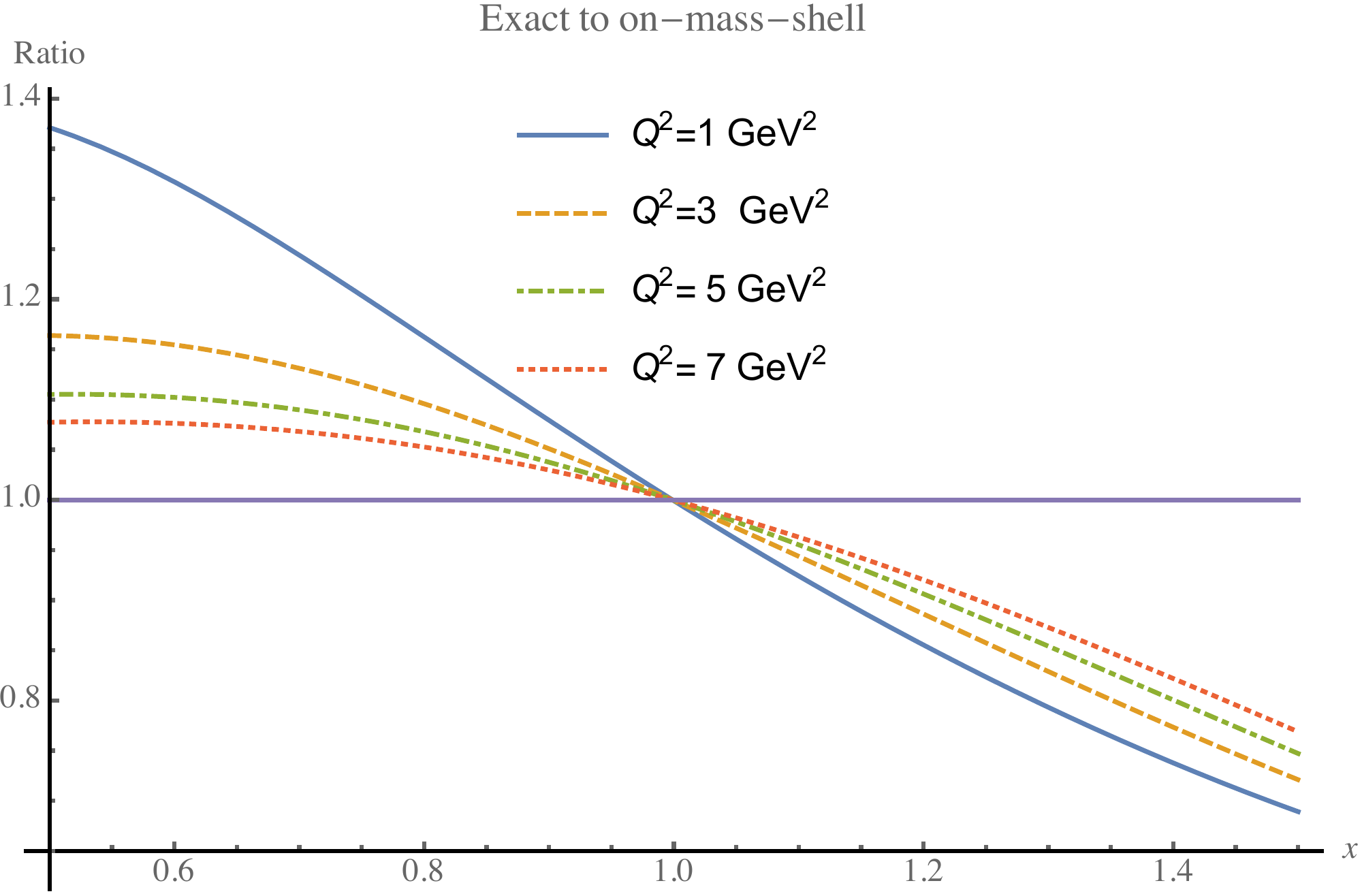}
\caption{(Color online)  Exact to on-mass shell ratios as a function of $x,\,\kp=0$ for different values of $Q^2$.}% Bottom: $Q^2$ from 100 to 600 GeV$^2$}
\end{figure}

%\end{document}
Fig.~6   shows  cross sections at $x=1$. We see that the exact and on-shell (0 virtuality) light-front approaches agree at all values of $Q^2$ that are shown. The curves for these two methods are not distinguishable. In contrast, the use of Bjorken scaling is not valid unless the value of $Q^2$ is very  high. 

That the accuracy of neglecting the virtuality holds only for values of $x$ near unity is shown in Fig.~7.  Significant errors are seen for values of $x$ lower and higher than unity. The accuracy improves as the value of $Q^2$ increases. However, these results show that the reliability of using light front wave functions is questionable if one is investigating high or low values of $x$ for  momentum transfers less than about 10 GeV$^2$.

All of the previous cross sections are obtained using relativistic wave functions. The non-relativistic approximation is studied in Fig.~8.  Using the non-relativsitic approximation fails except for vales of $x$ near unity. The relative errors increase significantly with increasing $Q^2$.  This indicates that  using standard non-relativistic wave functions  to analyze quasi-elastic scattering from deuteron targets may introduce an uncontrolled systematic error.
\begin{figure}[h]\label{p55}
\centering
\includegraphics[scale = .35]{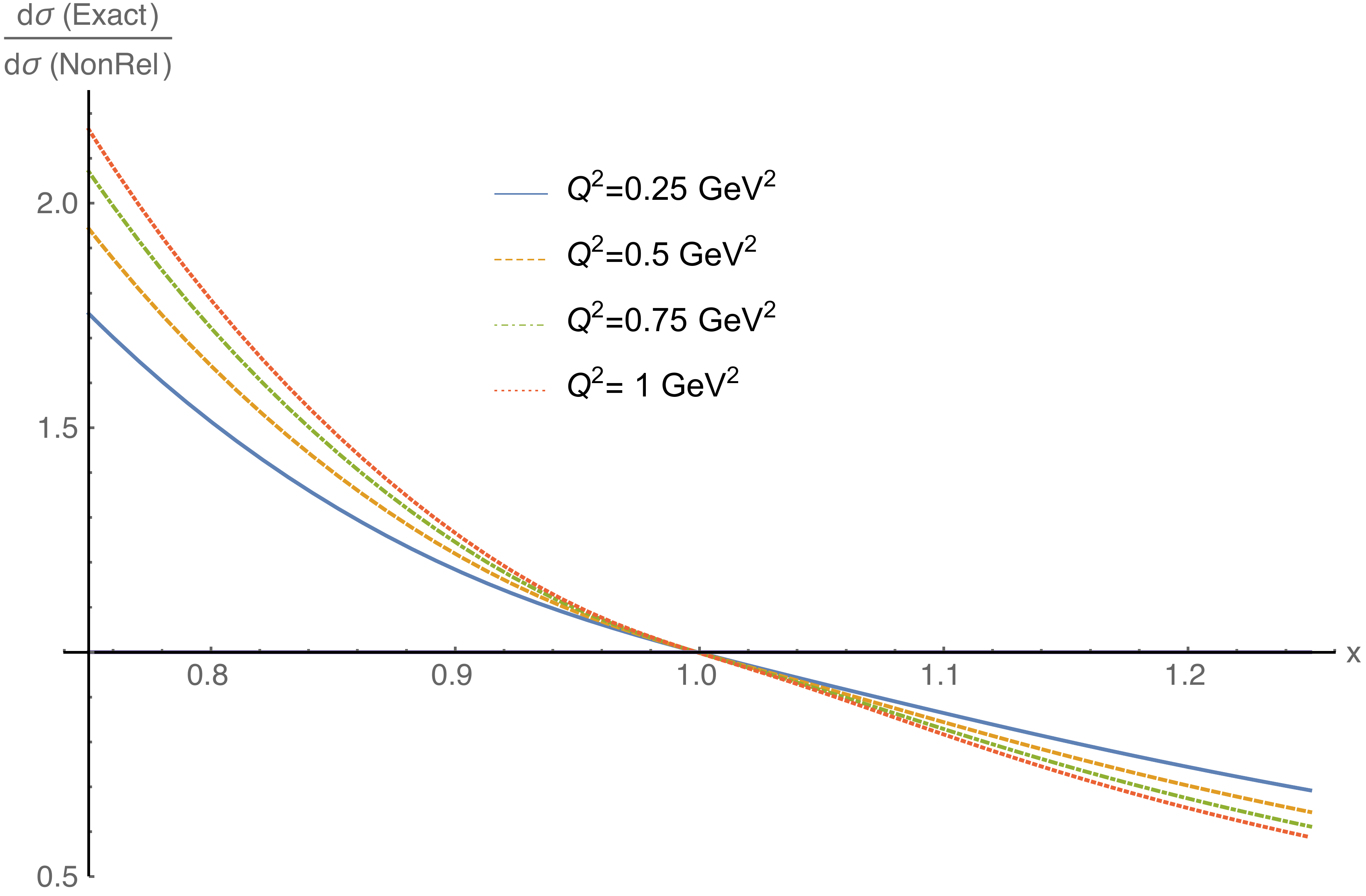}
\caption{(Color online) Non-relativistic approximation. Ratios of  exact to non-relativistic cross sections are shown  as a function of $x$ for different values of $Q^2$.}% Bottom: $Q^2$ from 100 to 600 GeV$^2$}
\end{figure}

\section{Other deuteron wave functions}

One might wonder if the qualitative results presented here are obtained only because of the simplicity of taking the deuteron vertex function to be a constant. Therefore we
 derive  and use  a more general wave function. Suppose instead of taking the vertex function to be constant $\G=G$, we postulate that, for example,
\bea \G(k,P)={G\, \L^2\over -k^2+\L^2+m^2},\eea
where $\L$ is a parameter to be determined. 
This means that the factor $1/X$ is replaced:
\bea {1\over X}={1\over k^2-m^2}\rightarrow {1\over k^2-m^2}{\L^2\over- k^2+\L^2+m^2} ={1\over k^2-m^2}-{1\over k^2-m^2-L^2}\equiv {1\over \tilde{X}}\label{other} \eea%-{1\over Y}\eea
The meaning of the second term may be identified by considering  the non-relativistic limit
 \bea
 {1\over \tilde{X}}=-{m\over M}\left[{1\over mB +\vec{k}^2}-{1\over mB  +\L^2{m\over M}+\vec{k}^2}\right],
\eea
and we observe that 
 $ {1\over \tilde{X}}$  is the Fourier transform of the Hulth\'en wave function~\cite{H1951} $(e^{-a r}-e^{-b r})/r$ with the parameters~\cite{Wong:1994sy}
 \bea a=\sqrt{Bm}=0.2316 \,{\rm fm}^{-1},\,b^2-a^2=\L^2m/M,\,b=1.3802\,{\rm fm}^{-1}.\eea
Evaluation yields $\L= 0.3795 $ GeV. 

The result \eq{other} provides an alternate model wave function, which can be treated using the four different wave functions discussed above.
\begin{figure}[h]\label{dse1}
\centering
\includegraphics[scale = .5]{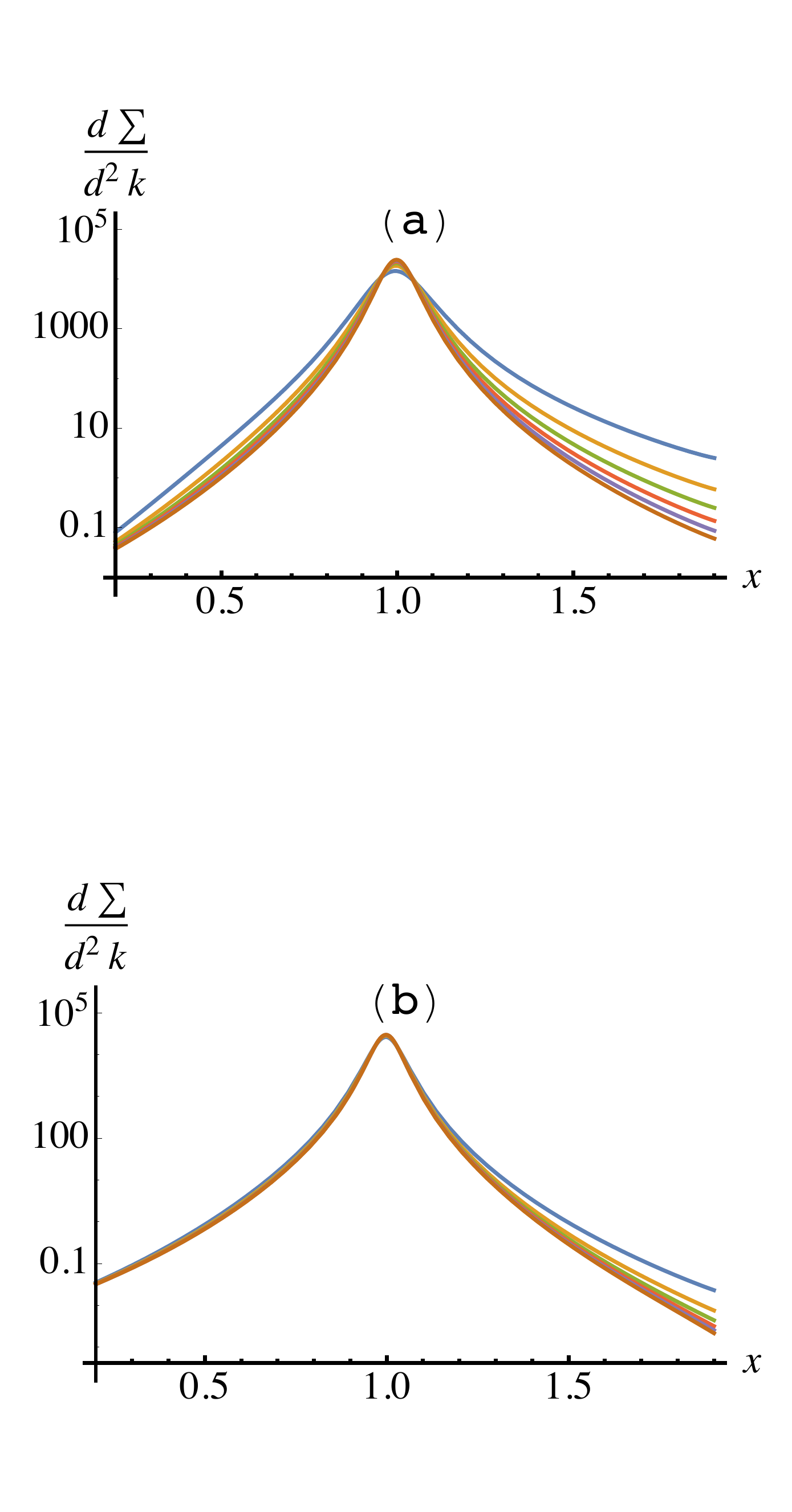}
\caption{(Color online) ${d\S\over  d^2\kp}$ as a function of $x$ for two different ranges of $Q^2$ with  $\kp=0$. (a) $Q^2$ from 1 to 6 GeV$^2$. (b)  $Q^2$ from 10 to 60 GeV$^2$.  For each case, the lower the value of $Q^2$, the higher the  cross section. For larger values of $Q^2$ the curves tend to coalesce.  Deuteron wave function of \eq{other}}% Bottom: $Q^2$ from 100 to 600 GeV$^2$}
\end{figure}

The use of this wave function is shown in Fig.~9. 
We see that the general shape and cross sections are about the  same as obtained using the wave function of Sect.~II. In particular, the requirement that $Q^2$ values of 10's of GeV$^2$ is reached to achieve scaling again occurs. 
 Fig.~10 shows again that  the differences between using  the exact spectator wave function instead of the light-front wave function are very substantial. Similarly, the non-relativistic approximation fails, see Fig.~11.   Thus the large effects of virtuality  shown in the previous section seem to be  general.

\begin{figure}[h]\label{p501}
\centering
\includegraphics[scale = .495]{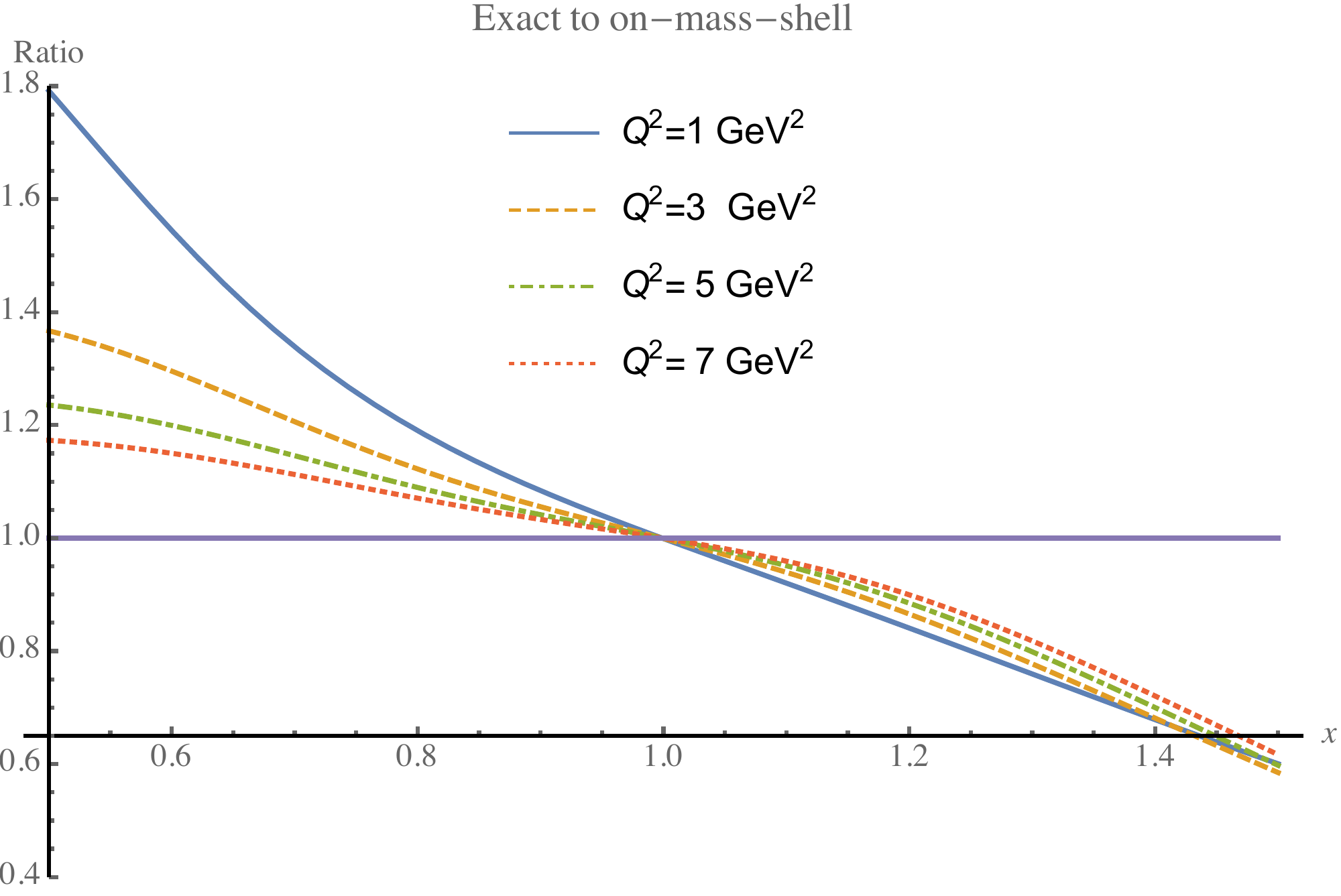}
\caption{(Color online)  Exact to on-mass shell ratios as a function of $x,\,\kp=0$ for different values of $Q^2$. Deuteron wave function of \eq{other}}% Bottom: $Q^2$ from 100 to 600 GeV$^2$}
\end{figure}
\begin{figure}[h]\label{p550}
\centering
\includegraphics[scale = .35]{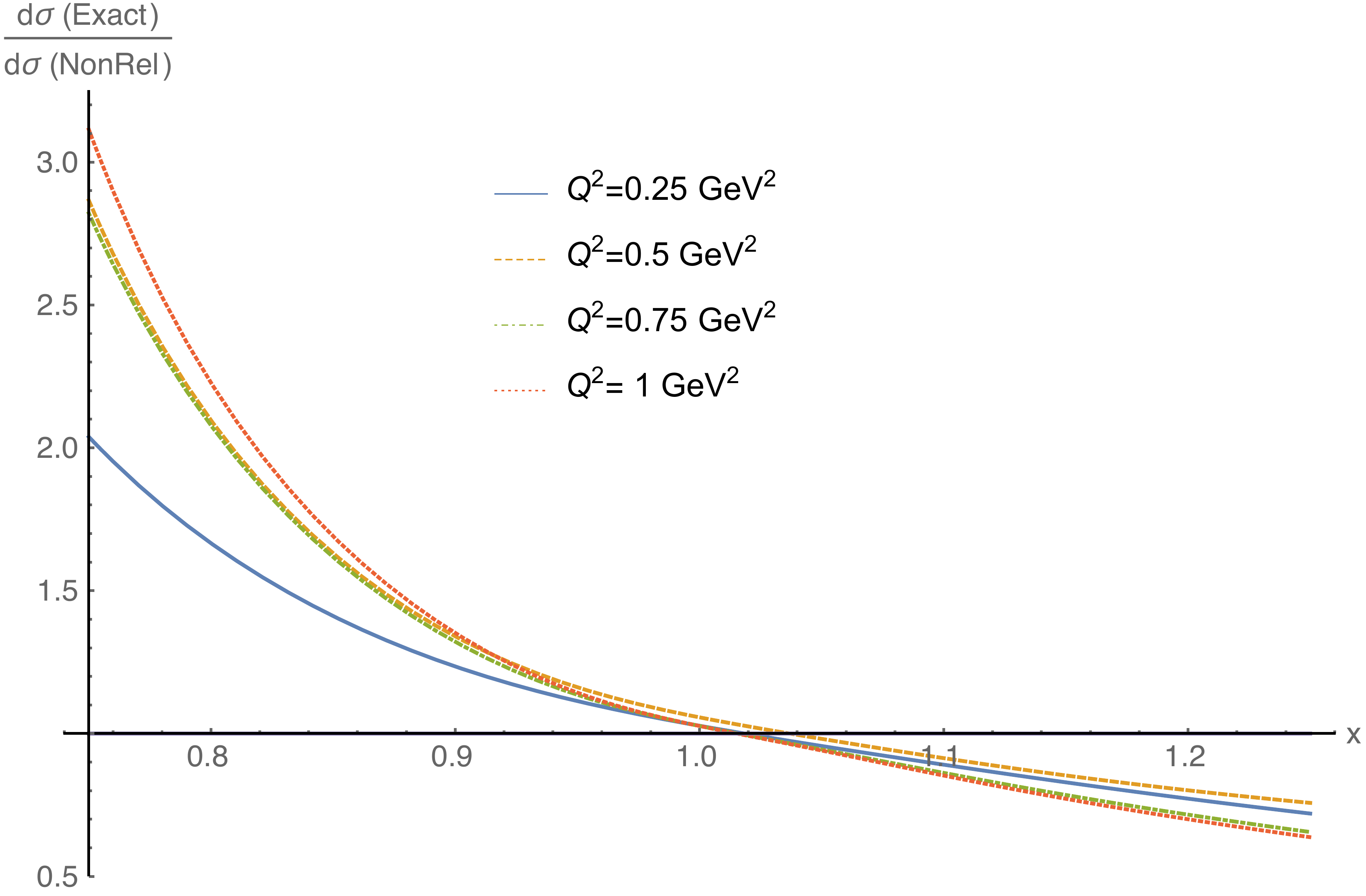}
\caption{(Color online) Non-relativistic approximation. Ratios of  exact to non-relativistic cross sections are shown  as a function of $x$ for different values of $Q^2$. Deuteron wave function of \eq{other}}% Bottom: $Q^2$ from 100 to 600 GeV$^2$}
\end{figure}

A final remark is that the model of \eq{other} can be generalized to match to any s-wave function in the non-relativistic limit.

\section{Virtuality of spin 1/2 fermions}
Previous sections used a simple model involving spin-less particles. In this section, we consider what happens when the virtual particle is a fermion. There is a lore stating that when 
evaluating matrix elements of so-called  ``good currents", with matrix elements  that go to infinity in the  infinite momentum frame, that the struck particles may be regarded as being on shell.
This lore is not generally correct, as we shall show.  The Feynman propagator of a virtual fermion of four momentum $p$ can be written as
\bea
&{1\over \fs{p} -m +i\e}= {(\fs{p}\,+m)\over p^2-m^2+i\e}={\sum_s u(p,s){\bar u}(p,s)\over p^2-m^2+i\e}+{\gamma^+\over 2p^+},\,\,p^+>0\label{y1}\\
&=-{\sum_s v(p,s){\bar v}(-p,s)\over p^2-m^2+i\e}+{\gamma^+\over 2p^+},\,\,p^+<0.\label{y2}
\eea 
In our notation the good current contains the operator  $\g^+$, and ${\g^+}^2=0$, so the second terms of \eq{y1} and \eq{y2} proportional to $\g^+$ do not contribute. Then one may use only on-shell spinors. This is the origin of the lore. However, there are two kinds of on-shell spinors, depending on whether the energy is positive or negative. For  a virtual particle the value of  $p^+$ can be positive or negative,  so that one may not neglect the possibility of  intermediate negative energy states having a significant influence.

\section{Summary and  Discussion}
Presently there is a strong   need for models of the deuteron wave function that are both realistic and relativistic. This need is driven by several experiments that aim at either determining deuteron structure or using known deuteron wave functions to determine neutron structure. In this paper, simple models are used to  show that   applying commonly used reductions of the Bethe-Salpeter equation from four dimensions to three dimensions severely compromises the ability to compute accurate cross sections for the interesting kinematic region in which the Bjorken $x$ variable differs from unity.   The only exact approach involves using the light-front-spectator wave function. 
In this case the wave function consists of one virtual constituent and one on-shell constituent.
There is one current experiment~\cite{Long:2014fda}  planned to specifically test the use of light-front spectator versus light-front wave functions. The considerations presented here encourage us to predict that only the light-front spectator wave functions would reproduce the experimental results.
 The present results indicate that the accurate interpretation of future experiments require the development of realistic relativistic light-front spectator nuclear wave functions.
\section*{Acknowledgments}
We thank C. Weiss and L. Weinstein for useful discussions. The work of JDT was supported by the National Science Foundation grant   PHY-1262811 as part of its REU program.
The work of  GAM was supported by the U.S. Department of Energy Office of Science, Office of Basic Energy Sciences program under Award Number DE-FG02-97ER-41014.

 \end{document}